\documentclass[pre,floats,aps,showpacs,amsmath,amssymb]{revtex4}
\usepackage{graphicx}
\usepackage{bm}

\begin{document}

\title{Mode-coupling approach to non-Newtonian Hele-Shaw flow}

\author{Magdalena Constantin\footnote{Present address: Department of Physics, 
Theoretical Condensed Matter Center,\\
University of Maryland, College Park, MD 20742} and Michael Widom}
\affiliation{Department of Physics, Carnegie Mellon University, 
Pittsburgh, PA  15213}

\author{Jos\'e A. Miranda}
\affiliation{Laborat\'{o}rio de F\'{\i}sica Te\'{o}rica e Computacional,
Departamento de F\'{\i}sica,\\ Universidade Federal de Pernambuco,
Recife, PE  50670-901 Brazil}

\begin{abstract}
The Saffman-Taylor viscous fingering problem is investigated for the
displacement of a non-Newtonian fluid by a Newtonian one in a radial
Hele-Shaw cell. We execute a mode-coupling approach to the problem and
examine the morphology of the fluid-fluid interface in the weak shear
limit. A differential equation describing the early nonlinear
evolution of the interface modes is derived in detail. Owing to
vorticity arising from our modified Darcy's law, we introduce a vector
potential for the velocity in contrast to the conventional scalar
potential.  Our analytical results address how mode-coupling dynamics
relates to tip-splitting and side branching in both shear thinning and
shear thickening cases. The development of non-Newtonian interfacial
patterns in rectangular Hele-Shaw cells is also analyzed. 
\end{abstract}


\maketitle

\section{Introduction}
\label{intro}

The celebrated Saffman-Taylor instability~\cite{MW_3} arises when a less 
viscous fluid pushes a more viscous one in the thin gap of a Hele-Shaw cell. 
The less viscous fluid can be either injected at an end of a channel-shaped 
cell (rectangular geometry)~\cite{MW_3}, or from the center of the cell 
(radial geometry)~\cite{Pat}. In both geometries the interface 
separating the fluids may deform, leading 
to the formation of fingerlike patterns, commonly known as viscous fingers. 
During the last four decades, the viscous fingering instability has been 
extensively studied, both theoretically~\cite{MW_1} and 
experimentally~\cite{MW_1b}. Much of the research in this area has examined 
the case in which the fluids involved are Newtonian. For Newtonian fluids, it 
is observed that the fingers grow and compete dynamically, resulting in 
a single stable finger in the rectangular geometry, and in patterned 
structures markedly characterized by the spreading, and subsequent 
splitting of the finger tips in the radial setup.

A whole different class of interfacial patterns arise 
when the Saffman-Taylor instability is studied by taking 
the displaced fluid as non-Newtonian~\cite{MW_1b,Van}. In contrast to most 
Newtonian fluids, non-Newtonian fluids differ widely in their 
hydrodynamic properties, with different fluids exhibiting a range of effects 
from elasticity and plasticity to shear thinning and shear 
thickening. Experiments using non-Newtonian 
fluids in radial and rectangular Hele-Shaw cells have revealed a wide 
variety of new patterns, showing snowflake-like 
shapes~\cite{MW_1b,Buka} and fracturelike structures~\cite{Lemaire,Zhao}.  
Instead of the traditional, tip-splitting-dominated Newtonian patterns, 
these experiments~\cite{MW_1b,Buka,Lemaire,Zhao} exhibit dendritic 
fingers and side branching. This morphological diversity and rich 
dynamical behavior motivated a number of theoretical 
studies of the problem~\cite{Sader,Bonn,Kondic_96,Kondic_98,Kondic_long}. 
One major difficulty faced by researchers is that, as opposed to 
the Newtonian case, the pressure field is no longer Laplacian. This 
implies a serious theoretical challenge, since {\it a priori} one would 
not be allowed to directly apply a Darcy's law approach 
to attack the problem.

The instability of radial Hele-Shaw flows involving non-Newtonian fluids has 
been studied theoretically by Sader {\it et al.}~\cite{Sader}. They considered 
power law fluids, and performed a linear stability analysis 
without making use of Darcy's law. Essentially, they showed that decreasing 
the power law index dramatically increases the growth rates leading 
to a more rapid development of the fingering patterns. 
A Darcy's law-type approach has been proposed by Bonn and 
collaborators~\cite{Bonn}. Bonn {\it et al.} suggested a 
modified Darcy's law including a shear rate dependent 
viscosity, and showed that, within their approach, the pressure 
field remains Laplacian. In a series of interesting papers Kondic {\it et 
al.}~\cite{Kondic_96,Kondic_98}, and subsequently Fast {\it et 
al.}~\cite{Kondic_long} extended reference~\cite{Bonn} ideas and 
derived a generalized Darcy's law from first principles, 
where viscosity depends upon the squared pressure gradient. 
It turns out that the Darcy's law formula 
proposed by Bonn {\it et al.}~\cite{Bonn} follows from the more basic 
version rigorously derived by Kondic {\it et al.}~\cite{Kondic_96,Kondic_98}. 
Efficient numerical simulations performed in 
references~\cite{Kondic_98,Kondic_long} have shown that shear 
thinning can suppress tip-splitting and leads to the formation of 
dendritic structures, presenting a clear side branching behavior.

Theoretical studies of the fully nonlinear stages of Hele-Shaw flow 
with non-Newtonian fluids rely heavily on intensive numerical 
simulations~\cite{Kondic_98,Kondic_long}. On the analytical 
side, the structure of the fingering dynamics in such complex fluids 
is largely restricted to linear stability 
investigations~\cite{Sader,Kondic_96}. Much less attention has been paid to 
the analytical investigation of the dynamics that bridges 
the {\it initial} (purely linear) and {\it final} (fully nonlinear) time 
regimes. In addition, theoretical as well as experimental analyses of 
flow of shear thickening fluids in Hele-Shaw cells still need to be 
addressed. In this paper we carry out the analytical weakly nonlinear analysis 
for the {\it intermediate} stages of evolution, and examine {\it both} 
shear-thinning and shear thickening cases. We adapt a weakly nonlinear 
approach originally developed to study Newtonian Hele-Shaw 
flows~\cite{MW_rectang,MW_rad}, to the non-Newtonian situation. 
We focus on the onset of the nonlinear effects, and try to understand 
how mode-coupling dynamics leads to basic morphological features and 
behaviors observed in non-Newtonian Hele-Shaw flows.

The article is organized as follows: section~\ref{derivation}
formulates our theoretical approach. We perform a Fourier
decomposition of the interface shape, and from an alternative form of
Darcy's law study the influence of weak shear effects on the
development of interfacial patterns. In contrast to the analysis for
Newtonian fluids, conventionally based on a scalar velocity potential,
we employ a vector velocity potential capable of describing vorticity
arising from the non-Newtonian fluid flow. Coupled, nonlinear,
ordinary differential equations governing the time evolution of
Fourier amplitudes are derived in detail. Section~\ref{discussion}
discusses both linear and weakly nonlinear dynamics. It concentrates
on the effect of shear thinning and shear thickening on finger
tip-splitting and side branching.  Section~\ref{1} briefly discusses
our linear stability results. Linear results are useful and
instructive, but do not allow accurate predictions of important
interfacial features. In section~\ref{2nd} we show that some of such
features can indeed be predicted and more quantitatively explained by
our analytical mode-coupling approach. At second order we find a
mechanism for finger tip-splitting in non-Newtonian Hele-Shaw flow: it
is suppressed (favored) for shear thinning (thickening) fluids.  Our
results indicate absence of side branching in the weak shear limit and
early flow stages, but suggest that it could be enhanced (inhibited)
for shear thinning (thickening) fluids. Section~\ref{rectangular}
discusses mode-coupling in rectangular flow geometry. Our chief
conclusions are summarized in Sec.~\ref{conclude}.

\begin{figure}[htbp]
  \begin{center}
    {\resizebox*{8.5cm}{!}{\includegraphics{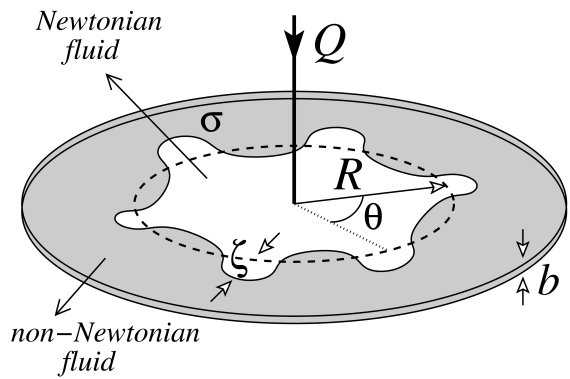}}}
      \caption{Schematic configuration of radial flow in a Hele-Shaw cell. The 
inner fluid is Newtonian and has negligible viscosity. The outer fluid is 
non-Newtonian. All physical parameters are defined in the text.}
    \label{fig:1}
  \end{center}
\end{figure}

\section{The mode-coupling differential equation} 
\label{derivation}

The Hele-Shaw cell (Fig.1) consists of two parallel plates separated by a 
small distance. The cell thickness 
$b$ is considered to be much smaller that any other length scale in 
the problem, so that the system is essentially two-dimensional. 
Consider the displacement of a viscous, non-Newtonian fluid by a Newtonian 
fluid of negligible viscosity in such confined geometry. The surface tension 
between the fluids is denoted by 
$\sigma$. The Newtonian fluid is injected at a constant areal flow rate $Q$ 
at the center of the cell.

The initial circular fluid-fluid interface is slightly perturbed, 
${\cal R } = R(t) + \zeta \left(\theta,t \right)$ ($\zeta/R \ll 1$), 
where the time dependent unperturbed radius is given by 
\begin{equation}
\label{R}
R(t) = \sqrt {R_{0}^{2}+\frac{Q t}{\pi}},
\end{equation}
$R_{0}$ being the unperturbed radius at $t=0$. The interface perturbation is 
written in the form of a Fourier expansion
\begin{equation}
\label{z}
\zeta(\theta,t)=\sum_{n=-\infty}^{+\infty} \zeta_{n}(t) \exp{\left(i n \theta 
\right)}, 
\end{equation}
where $\zeta_{n}(t) = (1/2\pi) \int_{0}^{2\pi} \zeta(\theta,t) 
\exp{(-in\theta)} ~{\rm d}\theta$ denotes the complex Fourier mode amplitudes 
and $n$=0, $\pm 1$, $\pm2$, $...$ is the discrete polar wave number. In our 
Fourier expansion~(\ref{z}) we include the $n=0$ mode to keep the area of the 
perturbed shape independent of the perturbation $\zeta$. Mass conservation 
imposes that the zeroth mode is written in terms of the other modes as
$\zeta_{0}= -(1/2R) \sum\limits_{n \ne 0} |\zeta_{n}(t)|^{2}$.

The relevant hydrodynamic equation for Newtonian Hele-Shaw flows, Darcy's 
law~\cite{MW_3,MW_1}, states that the gradient of the pressure is 
proportional to the fluid velocity, and oriented in the opposite 
direction with respect to the fluid flow
\begin{equation}
\label{Darcy}
{\nabla} p = - \frac{12\mu}{b^{2}}~{\bf v},
\end{equation}
where ${\bf v}={\bf v}(r,\theta)$ and $p=p(r,\theta)$ are, 
respectively, the gap-averaged velocity and pressure in the fluid. The 
viscosity of the fluid is represented by $\mu$. To model non-Newtonian 
Hele-Shaw flows, we use a suitable form of the Darcy's law in the 
weak shear thinning or thickening limit. We follow Bonn {\it et 
al.}~\cite{Bonn} and consider their shear rate dependent viscosity 
in the weak non-Newtonian limit $(\omega \tau)^{2} \ll 1$
\begin{equation}
\label{approx_visc}
{\mu(\omega^{2})}\approx {\mu_{0} \left[ 1-\left(1-\alpha\right) \omega^{2} 
\tau^{2} \right ]},
\end{equation}
where $\omega \approx v/b$ is the shear rate, $v=|{\bf v}|$, 
$\tau$ denotes the characteristic relaxation time of 
the fluid, and $\mu_{0}$ is a constant, 
zero-shear viscosity. The parameter $\alpha$ measures the shear dependence: 
$\alpha=1$ corresponds to the Newtonian fluids, and $\alpha<1$ ($\alpha>1$) 
gives the shear thinning (thickening) case. 

By substituing Eq.~(\ref{approx_visc}) into Eq.~(\ref{Darcy}) 
we obtain an alternative form of Darcy's law ideally suited to describe 
weak non-Newtonian effects
\begin{equation}
\label{new_Darcy}
{\nabla P} = - {\bf v} ~+~ \delta v^{2} {\bf v}.
\end{equation}
Here $P=[b^{2}/(12 \mu_{0})] ~p$ represents a generalized pressure field, and 
$\delta=(1-\alpha)(\tau/b)^{2}$ is a small parameter that expresses the 
non-Newtonian nature of the displaced fluid: $\delta=0$ corresponds to the 
Newtonian case, while $\delta>0$ ($\delta<0$) describes the shear thinning 
(thickening) case. Taking the divergence
of Eq.~(\ref{new_Darcy}) and using incompressibility ($\nabla \cdot
{\bf v}=0$) the pressure is seen to be anharmonic (nonvanishing
Laplacian). Our subsequent analysis incorporates this effect, in
contrast to the treatment in Ref.~\cite{Bonn}.  Indeed, an alternate
route to our Eq.~(\ref{new_Darcy}) is to start with viscosity depending
on the square of the pressure gradient (as in Kondic, {\it et
al.}~\cite{Kondic_96}), then approximate the pressure gradient with the
velocity for small $\delta$.

Our pertubative approach keeps terms up to the second order 
in $\zeta$ and up to first order in $\delta$. Considering the fact that the 
viscosity of the Newtonian fluid is negligible, the generalized pressure jump 
condition at the interface can be written as~\cite{MW_1}
\begin{equation}
\label{p_jump}
P|_{{\cal R}} =-\gamma ~\kappa_{\|}|_{{\cal R}},
\end{equation}
where $\gamma= b^{2} \sigma/(12 \mu_{0})$ and $\kappa_{\|}$ is the 
curvature in the direction parallel to the plates.

The weakly nonlinear approach to radial, Newtonian Hele-Shaw flow
developed in reference~\cite{MW_rad}, related the fluid velocity to a
{\it scalar} velocity potential ${\bf v}=-\nabla \phi$, this
replacement made possible by the irrotational nature of the flow for
Newtonian fluids. For non-Newtonian fluids, in contrast, flows
governed by the modified Darcy's law (\ref{new_Darcy}) exhibit
vorticity. Hence we perform our calculations
using a {\it vector} potential ${\bf v}= \nabla
\times {\bf A}$. The most general form of the vector potential 
can be written as
\begin{equation}
\label{gen_vec_pot}
{\bf A}= \left[ \frac{Q}{2\pi} \theta+ \sum_{m,n\not=0} A_{mn} 
\left(\frac{R}{r} \right)^{m} \exp{(i n \theta)} \right] \hat {\bf z},
\end{equation}
where $A_{mn}$ are the Fourier coefficients of the velocity vector potential 
and $\hat {\bf z}$ is the outward unit-normal to the upper cell plate. The 
radial and polar components of the fluids velocities are
\begin{equation}
\label{gen_vel}
{v_{r}}=\frac{Q}{2 \pi r}+\sum_{m,n\not=0} i n A_{mn} 
\left(\frac{R^{m}}{r^{m+1}} \right)\exp{(i n \theta)},
\end{equation}
and
\begin{equation}
\label{general_vel}
{v_{\theta}}=\sum_{m,n\not=0} m A_{mn} \left(\frac{R^{m}}{r^{m+1}} 
\right)\exp{(i n \theta)}.
\end{equation}
Note that the vector potential reduces to the unperturbed 
steady flow with a circular interface ($v_r=Q/2\pi r$, $v_{\theta}=0$)
when $R\rightarrow 0$ and also when $r\rightarrow \infty$.

We exploit the fact that $\nabla P$ must be curl free, and 
impose the so-called solvability condition $ \nabla \times  \nabla 
P=0$. It simplifies the general form of the vector potential expansion given 
in Eq.~(\ref{gen_vec_pot}). The solvability condition reveals that, without 
loss of generality, one can rewrite the vector 
potential as
\begin{equation}
\label{vec_pot}
{\bf A}= \left \{ \frac{Q}{2\pi} \theta+ \sum_{n\not=0} 
A_{n}\left(\frac{R}{r} \right)^{|n|} 
\exp{ \left(i n \theta \right)} + \delta \left[ \sum_{n\not=0} B_{n} \left( 
\frac{R}{r} 
\right)^{|n|} \frac{1}{r^{2}} \exp{(i n \theta)} \right] \right \}~\hat {\bf 
z},
\end{equation}
replacing the array of coefficients $A_{mn}$ with the simpler set of
$A_n$ and $B_n$.  Observe that the vector potential~(\ref{vec_pot}) is
simply a superposition of a purely Newtonian term ($\propto \delta^{0}$,
coefficients $A_n$) and a non-Newtonian contribution ($\propto \delta^{1}$,
coefficients $B_n$)
\begin{equation}
\label{vec_pot_formal}
{\bf A}= {\bf A}_{N}+ {\bf A}_{NN}.
\end{equation}
The flow described by ${\bf A}_N$ is irrotational, while ${\bf A}_{NN}$
has a curl.

Similarly, we express the pressure of the outer fluid as a sum of Newtonian 
and non-Newtonian pressures, and propose a general form for their Fourier 
expansion
\begin{equation}
\label{pressure_formal}
P= P_{N}+ P_{NN},
\end{equation}
where
\begin{equation}
\label{pressure_formal1}
P_{N}= -\frac{Q}{2\pi} \log{\left(\frac{r}{R}\right)}+ \sum_{n\not=0} p_{n} 
\left(\frac{R}{r} \right)^{|n|}\exp{(i n \theta)},
\end{equation}
and
\begin{equation}
\label{pressure_formal2}
P_{NN}= \delta \left[-\left(\frac{Q}{2\pi} \right)^{3} \frac {1}{2r^{2}} + 
\sum_{n\not=0} q_{n} \left(\frac{R}{r} \right)^{|n|} \frac {1}{r^{2}}\exp{(i 
n \theta)} \right].
\end{equation}

The gradient of the complex pressure field~(\ref{pressure_formal}) 
must satisfy the non-Newtonian  Darcy's law 
given by Eq.~(\ref{new_Darcy}). By inspecting 
the $r$ and $\theta$ components of~(\ref{new_Darcy}), and by 
examining the Newtonian and non-Newtonian components of it, we 
can express the Fourier coefficients of 
$P_{N}$, $P_{NN}$ and ${\bf A}_{NN}$ in terms of the Fourier coefficients of 
${\bf {A}}_{N}$, 

\begin{equation}
\label{p_n}
p_{n}= {\rm sgn}(n) \left( i A_{n}\right),
\end{equation}

\begin{equation}
\label{q_n}
q_{n}= - i A_{n} \beta(n) \frac{n}{|n|} \left(\frac{Q}{2\pi}\right)^{2}
+\frac{Q}{4 \pi \left(|n|+1 \right)} \sum_{m\not=0,m\not=n} m A_{m} 
\left (n-m \right) A_{n-m} ~k\left (n, m \right),
\end{equation}

\begin{equation}
\label{B_n}
B_{n}=  -A_{n} \alpha(n) \left(\frac{Q}{2\pi}\right)^{2} +\frac{Q}{4 \pi 
\left(|n|+1 \right)}
\sum_{m\not=0,m\not=n} m \left(i A_{m} \right) \left (n-m \right) A_{n-m} 
~h\left (n, m \right),
\end{equation}

where in order to keep the results in a compact form, we introduced the 
coefficients
\begin{equation}
\label{alpha}
{\alpha \left(n \right)} = { \frac{|n| \left(|n|-1 
\right)}{2\left(|n|+1\right)}},
\end{equation}

\begin{equation}
\label{beta}
{\beta \left(n \right)}= { \frac{|n| \left(|n|+3 
\right)}{2\left(|n|+1\right)}},
\end{equation}

\begin{equation}
\label{h}
h \left(n, m \right)=\left(|n|+2 \right) {\rm sgn}(n-m) - \frac{n}{2} \left \{ 
3 - 
{\rm sgn} \left[m \left(n-m \right) \right]\right \},
\end{equation}
and
\begin{equation}
\label{k}
k \left(n, m \right)= -n \,{\rm sgn}(n-m) + \frac{1}{2} \left(|n|+2 \right) 
\left \{3-{\rm sgn} 
\left[m \left(n-m \right)\right]\right \}.
\end{equation}
Note that ${\rm sgn}(n)=1$ if $n>0$ and ${\rm sgn}(n)=-1$ if $n<0$. 

Using Eqs.~(\ref{p_n})-(\ref{B_n}), which are consistent with the solvability 
condition and Darcy's law~(\ref{new_Darcy}), we can derive the general 
expression of the vector potential Fourier coefficients in terms of the 
perturbation amplitudes. To fulfill this goal, consider the kinematic boundary 
condition~\cite{MW_1,MW_1b} 
\begin{equation}
\label{kinem_b_c}
{\frac{\partial \cal R}{\partial t}}= \left[ \frac{1}{r} \frac{\partial \cal 
R}{\partial \theta} 
\left(-v_{\theta} \right)  + v_{r} \right]_{|_{\cal R}},
\end{equation}
which states that the normal components of each fluid's velocity at the 
interface equals the velocity of the interface itself. 
By expanding Eq.~(\ref{kinem_b_c}) up to the 
second order in $\zeta$ and up to first order in $\delta$ we find the 
coefficient of the vector potential corresponding to the $n$-th evolution 
mode, $A_{n}^{(k)}$, in terms of $\delta$ and the $k$-th order in $\zeta$ 
($k$ = 1, 2)
\begin{equation}
\label{vec_pot1}
{i A_{n}^{\left(1\right)}(t)}=\left[\frac{R} {n} \dot \zeta_{n}+\frac{\dot R} 
{n} 
\zeta_{n} \right] \left[ 1+ \delta {\dot R}^{2} \alpha \left(n \right) 
\right], 
\nonumber \\
\end{equation} 

\begin{eqnarray}
\label{vec_pot2}
{i A_{n}^{\left(2\right)}(t)} & = & \frac{{\dot R}}{R}\sum_{m \not=0,n} \left 
[\frac 
{|m|} {m} + \delta {\dot R}^{2} u(n,m)\right ] \zeta_{m}\zeta_{n - m} 
 + \sum_{m \neq 0,n} \left [ \frac{1}{n}+\frac {|m|} {m} 
+ \delta {\dot R}^{2} v(n,m) \right ] \dot \zeta_{m}\zeta_{n - m} \nonumber \\ 
                              & - & \frac{\delta {\dot R} 
R}{2\left(|n|+1\right)} 
\sum_{m \neq 0,n} h(n,m) ~\dot \zeta_{m} \dot \zeta_{n - m}
-\frac{\delta {\dot R}^{2}}{2\left(|n|+1\right)}\sum_{m \neq 0,n} h(n,m) 
~\zeta_{m} 
\dot \zeta_{n - m}, \nonumber \\
\end{eqnarray}
where the overdot denotes total time derivative, ${\dot R}=Q/(2 \pi R)$ 
represents the unperturbed interface velocity, and the coefficients
\begin{equation}
\label{coeff_u}
{u \left(n, m \right)}=\alpha(n)\frac{|m|}{m}-2\frac{\alpha(m)}{m} 
-\frac{h(n,m)}{2\left(|n|+1 \right)},
\end{equation}

\begin{equation}
\label{coeff_v}
{v \left( n, m \right)}=\alpha(n)\left
(\frac {1}{n}+\frac{|m|}{m}\right)-2\frac{\alpha(m)}{m}
-\frac{h(n,m)}{2\left(|n|+1 \right)}.
\end{equation}

To conclude our derivation we need one more step. The vector 
potential coefficients can be introduced into the pressure jump condition 
(Eq.(\ref{p_jump})) and using Darcy's law (Eq.(\ref{new_Darcy})) one can 
finally find the equation of motion for perturbation amplitudes $\zeta_{n}$. 
We present the evolution of the perturbation amplitudes in terms of $\delta$ 
and the $k$-th order in the perturbation amplitude $\zeta$
\begin{equation}
\label{evol_eq}
{\dot \zeta_{n}}=\dot \zeta_{n}^{\left(1\right)} + \dot 
\zeta_{n}^{\left(2\right)},
\end{equation} 
where
\begin{equation}
\label{1_ord_eq}
{\dot \zeta_{n}^{\left(1\right)}}= \lambda \left(n\right) \zeta_{n},
\end{equation} 

\begin{equation}
\label{lambda}
{\lambda(n)}=\frac{\dot {R}}{R}\left(|n|-1\right)-\frac{\gamma}{R^{3}} |n| 
\left (n^{2}-1 \right) + \delta \dot {R}^{2} \frac{|n|}{R}\left[ \dot {R}
\frac{|n|-1}{|n|+1}
-\frac {2 \gamma}{R^{2}} \frac{|n| \left(n^{2}-1\right)}{|n|+1}\right],
\end{equation}
is the non-Newtonian linear growth rate first reported in Kondic, {\it et 
al.}~\cite{Kondic_96}, and

\begin{eqnarray}
\label{2_ord_eq}
{\dot \zeta_{n}^{\left(2\right)}} &=& \sum_{m \not=n,0} \left [F_{N}(n,m) + 
\delta  ~F_{NN}(n,m)\right ] 
~\zeta_{m}\zeta_{n - m} + \sum_{m \neq n,0} \left [G_{N}(n,m) + \delta 
~G_{NN}(n,m) \right ] ~\dot \zeta_{m}\zeta_{n - m} \nonumber\\ 
                                  &+& \delta \sum_{m \neq n,0} H_{NN} 
\left(n,m \right) ~\zeta_{m} \dot \zeta_{n - m} ~+~ \delta \sum_{m \neq n,0} 
J_{NN}(n,m) ~\dot \zeta_{m} \dot \zeta_{n - m}.
\end{eqnarray} 
In Eq.~(\ref{2_ord_eq}) the coefficients $F_{N}, F_{NN}, G_{N}, G_{NN}, 
H_{NN}$, and $J_{NN}$ represent the second order Newtonian (N) and 
non-Newtonian (NN) terms. Their detailed functional form are presented 
in the appendix. An important feature of the second order coefficients 
is that they present special reflection symmetries
\begin{eqnarray}
\label{symm}
{{\cal C} \left(n,-m \right)}= {\cal C} \left(-n,m \right) \nonumber\\
{{\cal C} \left(-n,-m \right)}= {\cal C} \left(n,m \right)
\end{eqnarray}
where ${\cal C}=F_{N},F_{NN}, G_{N}, G_{NN}$, $H_{NN}$, and $J_{NN}$, 
respectively. Equation~(\ref{evol_eq}) 
is the mode-coupling equation of the non-Newtonian Saffman-Taylor 
problem in radial geometry. It gives us the time 
evolution of the perturbation amplitudes $\zeta_{n}$, accurate 
to second order, in the weak shear limit. The rest of the paper uses 
Eq.~(\ref{evol_eq}) to study the development 
of interfacial instabilities, and to examine how the non-Newtonian 
parameter $\delta$ affects pattern morphology.

\section{Discussion}
\label{discussion}

In the next three subsections we use our mode-coupling approach to investigate 
the interface evolution at first and second order. To 
simplify our discussion it is convenient to rewrite the net perturbation 
~(\ref{z}) 
in terms of cosine and sine modes
\begin{equation}
\label{sincos}
\zeta(\theta,t)= \zeta_{0} + 
\sum_{n = 1}^{\infty} \left[ a_{n}(t)\cos(n\theta) + b_{n}(t)\sin(n\theta) 
\right ],
\end{equation}
where $a_{n}=\zeta_{n} + \zeta_{-n}$ and $b_{n}=i \left ( \zeta_{n} -
\zeta_{-n} \right )$ are real-valued. Without loss of generality we
may choose the phase of the fundamental mode so that $a_{n} > 0$ and
$b_{n}=0$.

\subsection{First order}
\label{1}

Although at the level of linear analysis we do not expect to detect 
or rigorously predict important nonlinear effects 
such as tip-splitting and side branching, some useful information 
may still be extracted. A nice example of how purely linear results 
can help to understand complicated morphological features appearing in 
non-Newtonian radial Hele-Shaw flows is found in 
references~\cite{Kondic_96,Kondic_98}. By studying their linear growth rate, 
Kondic {\it et al.}~\cite{Kondic_96,Kondic_98} found that shear thinning 
decreases the wave number of maximal growth, 
increases the maximum growth rate, and tightens the band of unstable modes. 
Based on this increased selectivity of wavelengths, they 
postulated that shear thinning can lead to 
suppression of tip-splitting. Their speculations 
have not been further investigated analytically, but instead 
have been supported by their own intensive numerical 
simulations~\cite{Kondic_98,Kondic_long}.

\begin{figure}[htbp]
  \begin{center}
    {\resizebox*{9cm}{!}{\includegraphics{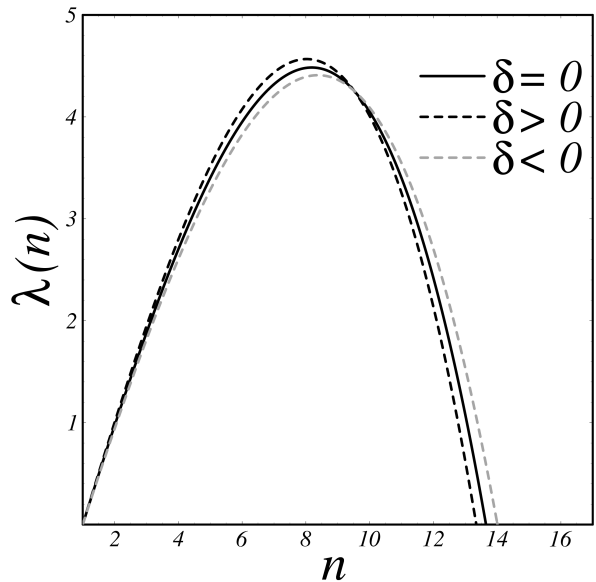}}}
      \caption{Linear growth rate~(\ref{lambda}) as a function of mode 
number $n$ for three different values of the non-Newtoniam parameter: 
$\delta=0$ (black solid curve), +0.05 (black dashed curve), -0.05 (gray dashed 
curve). The injection rate $Q=2 \pi ~{\rm cm^{2}/s}$, 
$R=1 ~{\rm cm}$, and $\gamma=1/200~{\rm cm^{3}/s}$. The units of 
$\lambda(n)$ and $\delta$ are ${\rm s}^{-1}$ and $({\rm cm/s})^{-2}$, 
respectively.}
    \label{fig:2}
  \end{center}
\end{figure}

Our first order results, namely our linear growth rate 
expression~(\ref{lambda}), agrees with Kondic {\it et 
al.}~\cite{Kondic_96} equivalent formula, if 
we set their constant $C$ equal to ${\dot{R}}^{2}$. We 
use growth rate~(\ref{lambda}) to gain insight 
into shear thinning/thickening behavior. Figure 2 plots 
$\lambda(n)$ as a function of mode 
number $n$ for three different values of the 
non-Newtoniam parameter: a) $\delta=0$; 
b) $\delta=0.05$; and c) $\delta=-0.05$. By inspecting Fig. 2, 
we notice that, unlike the shear thinning case ($\delta > 0$) discussed in 
references~\cite{Kondic_96,Kondic_98}, shear thickening ($\delta < 0$) 
widens the band of 
unstable modes, and decreases growth rate for the wave number of maximal 
growth. So, by applying similar arguments as those used by Kondic {\it et 
al.}~\cite{Kondic_96,Kondic_98} for the weak shear thinning case, 
we postulate that shear thickening would lead to {\it enhanced} 
tip-splitting. Unfortunately, for the shear thickening radial Hele-Shaw flow, 
both numerical simulations and experiments are 
not available in the literature to 
confirm this claim. Of course, it is also of interest to study such a 
possibility analytically. This is one of the topics we examine in 
Sec.~\ref{2nd}.

\subsection{Second order}
\label{2nd}

On the analytical side, one must go beyond linear analysis in order to 
investigate in more detail the main morphological features found in Hele-Shaw 
flows with both shear thinning and shear thickening fluids. To this day, 
besides a few first order linear growth 
studies~\cite{Kondic_96,Kondic_98,Kondic_long}, there are no other analytical 
results concerning non-Newtonian fluid evolution in radial 
Hele-Shaw geometry. If on one hand the non-Newtonian problem is vastly richer, 
it is also true it is much less amenable to analytic attack than its Newtonian 
counterpart. Our mode-coupling approach intends to provide a better and more 
complete analytical understanding of the complex non-Newtonian time evolution 
dynamics in Hele-Shaw cells.

\subsubsection{Action of the parameter $\delta$ on finger tip-splitting}
\label{2nd2}

We use the mode-coupling equation~(\ref{evol_eq}) to investigate the coupling 
of a small number of modes. At second order the most noteworthy 
effect refers to the action of the non-Newtonian parameter $\delta$ on finger 
tip-splitting.  Tip-splitting is related to the influence of a fundamental 
mode $n$ on the growth of its harmonic $2n$~\cite{MW_rad}. 
For consistent second order expressions, we replace 
the time derivative terms $\dot{a}_{n}$ and $\dot{b}_{n}$ by 
$\lambda(n)~a_{n}$ and $\lambda(n)~b_{n}$, respectively. Under these 
circumstances the equation of motion for the 
cosine mode $2n$ is
\begin{equation}
\label{2_harm}
\dot{a}_{2n}=\lambda(2n)~a_{2n} + \frac{1}{2} ~T(2n,n) ~a_{n}^2,
\end{equation}
where the function that multiplies $a_{n}^{2}$, $T(2n,n)$, is called the 
tip-splitting function and its general expression is  
\begin{eqnarray} 
\label{T(n,m)} 
T(n,m) & = & F_{N}(n, m) + \lambda(m)~G_{N}(n, m) + \delta ~[  
F_{NN}(n,m) + \lambda (m) ~G_{NN}(n,m) \nonumber\\ 
        & + & \lambda (n-m) ~H_{NN}(n,m) +  
\lambda(m) ~ \lambda (n-m) ~J_{NN}(n,m)]. 
\end{eqnarray} 
Equation~(\ref{2_harm}) shows that the presence of the fundamental 
mode $n$ forces growth of the harmonic mode $2n$. The function 
$T(2n,n)$ acts like a driving force and its sign dictates if finger 
tip-splitting is favored or not by the dynamics. If $T(2n,n)<0$, $a_{2n}$ 
is driven negative, precisely the sign that leads to finger tip-widening and 
finger tip-splitting. If $T(2n,n)>0$ growth of $a_{2n} > 0$ would be favored, 
leading to outwards-pointing finger tip-narrowing.

\begin{figure}[htbp]
  \begin{center}
    {\resizebox*{9cm}{!}{\includegraphics{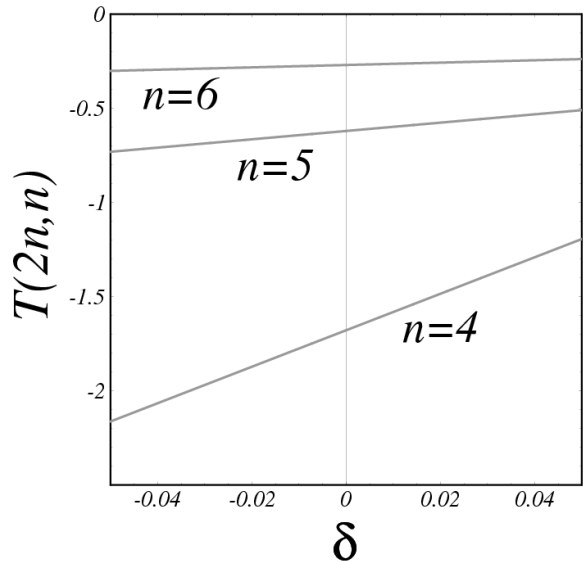}}}
      \caption{Variation of $T(2n,n)$ as a function of the non-Newtonian 
parameter 
$\delta$, for three different Fourier modes ($n=4,5,6$).
The injection rate $Q=2 \pi ~{\rm cm^{2}/s}$ and $\gamma=1/100~{\rm 
cm^{3}/s}$. The units of $T(2n,n)$ and $\delta$ are $({\rm cm~s})^{-1}$ and 
$({\rm cm/s})^{-2}$, respectively.}
    \label{fig:3}
  \end{center}
\end{figure}

To investigate the influence of the non-Newtonian parameter $\delta$ on 
tip-splitting behavior at second order, we plot in Fig. 3 the behavior of 
$T(2n,n)$ as a function of $\delta$, for a few Fourier modes ($n=4,5,6$). 
To simplify our analysis we adopt an instantaneous approach: we consider a 
particular ${\dot R}$ and $R$ combination, using the identity ${\dot R}R=Q/2 
\pi$, at the onset of growth of mode $2n$ [using the condition 
$\lambda(2n)=0$] in the Newtonian limit $\delta=0$, where we know $T(2n,n)$ is 
negative~\cite{MW_rad}. We see from Fig. 3 that the curve for mode $n=4$ lies 
below the curves associated to $n=5, 6$. This is an expected behavior 
since smaller values of $n$ would mean more room for the existing fingers to 
split. We also observe that the curves associated to smaller $n$ show a 
stronger angular inclination with respect to the horizontal $\delta$-axis. 
Therefore, lower Fourier modes would be more sensitive to variations in 
$\delta$.

Further inspection of Fig. 3 reveals that, for $\delta>0$, 
$T(2n,n)$ becomes less negative as $\delta$ increases, 
meaning that the interface has less tendency toward 
tip-splitting. In contrast, for $\delta<0$ we observe that 
$T(2n,n)$ becomes more negative as the magnitude of $\delta < 0$ 
increases, indicating an enhanced tendency of the fingers to split 
at their tips. We conclude that tip-splitting is suppressed for shear 
thinning fluids, and enhanced for shear thickening ones. This 
fact is more clearly illustrated in Fig. 4. In Fig. 4 we plot the 
fluid-fluid interface for a certain time ($t=30 ~{\rm s}$), considering 
the interaction of two cosine modes (a fundamental $n=4$ and its harmonic 
$2n=8$), for three diferrent values of $\delta$: (a) $\delta=0$ (solid curve);
(b) $\delta>0$ (black dashed curve); and (c) $\delta<0$ (gray dashed 
curve). From Fig. 4 it is evident that there is a stronger splitting in 
the shear thickening case.

These second order effects regarding tip-splitting are consistent with the
first order effects described in section~\ref{1}. With respect to shear 
thinning behavior, our analytical results agree with numerical 
simulations~\cite{Kondic_98,Kondic_long} and experiments~\cite{MW_1b,Buka} 
of fully nonlinear stages of interface evolution. In addition, we detect 
favored tip-splitting in the shear thickening case, a relevant 
nonlinear behavior not previously reported in the literature. We 
have also verified that, for a given mode $n$, smaller values of 
the surface tension parameter $\gamma$ lead to enhancement (suppression) 
of splitting events for $\delta < 0$ ($\delta > 0$).

\begin{figure}[htbp]
  \begin{center}
    {\resizebox*{9cm}{!}{\includegraphics{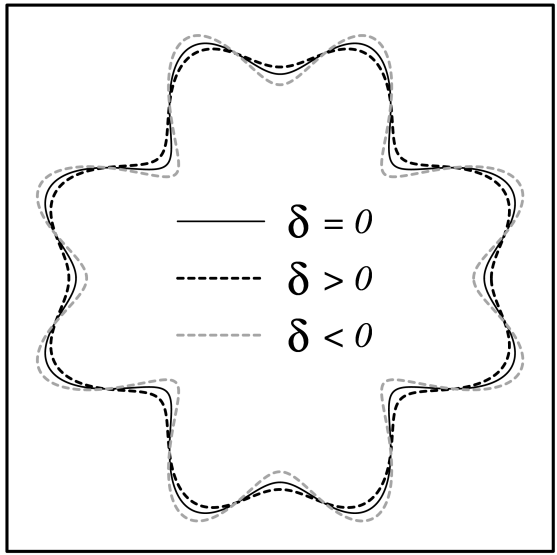}}}
      \caption{Snapshot of the fluid-fluid interface ($t=30~{\rm s}$), for the 
interaction of two cosine modes $n=4$ and $2n=8$. Three values of the 
non-Newtonian parameter are considered $\delta=0$ (black solid curve), 
$\delta=+0.01~({\rm cm/s})^{-2}$ (black dashed curve), and $\delta=-0.01~({\rm 
cm/s})^{-2}$ (gray dashed curve). Other physical parameters are: 
$a_{n}(0)=0.001~{\rm cm}$, $R_{0}=0.3 ~{\rm cm}$, $\gamma=0.025~{\rm 
cm^{3}/s}$, and $Q=3 \pi ~{\rm cm^{2}/s}$. Splitting is favored for 
shear-thickening fluids (gray dashed curve).}
    \label{fig:4}
  \end{center}
\end{figure}

A physical mechanism that seems to be at work in shear thinning case 
has been proposed in references~\cite{Kondic_96,Kondic_98,Kondic_long}:
viscosity is lowered in high-shear regions, which are in front of the fingers. 
Hence the tip of a finger experiences less resistance than the part of the 
interface near the tip. This suggests that tip-splitting would be suppressed 
if the fluid shear-thins. Our second order mode-coupling results indicate that 
a similar mechanism is at work in the shear thickening case: the resistance 
would be increased at the finger tips, suggesting that tip-splitting would be 
increased. 

\subsubsection{Action of the parameter $\delta$ on side branching}
\label{2nd3}

Another relevant non-Newtonian effect that can be studied at second
order refers to the side branching phenomenon. In the framework of a
mode-coupling theory, side branching requires the presence of mode
$3n$. If the harmonic mode $a_{3n}$ is positive and sufficiently
large, it can produce interfacial lobes branching out sidewards which
we interpret as side branching.

Consider the influence of the fundamental mode $n$, and its
harmonic $2n$, on the growth of mode $3n$. The equation of motion
for the cosine $3n$ mode is
\begin{equation}
\label{a_3n_ord2}
\dot{a}_{3n}=\lambda(3n)~a_{3n} + \frac{1}{2} ~S(3n)~a_{n} a_{2n},
\end{equation}
where the side branching function $S(3n)=\left[T(3n,n)+T(3n,2n)\right]$ 
can be easily obtained from Eq.~(\ref{T(n,m)}). By analyzing 
Eq.~(\ref{a_3n_ord2}) we observe that mode $3n$ can be 
spontaneously generated due to the 
driving term proportional to $a_{n} a_{2n}$, such that it enters
through the dynamics even when it is missing from the
initial conditions. The existence and phase of mode $3n$ depends on the 
interplay of the modes $n$ and $2n$. Side branching would be favored if 
$a_{3n}>0$.

\begin{figure}[htbp]
  \begin{center}
    {\resizebox*{9cm}{!}{\includegraphics{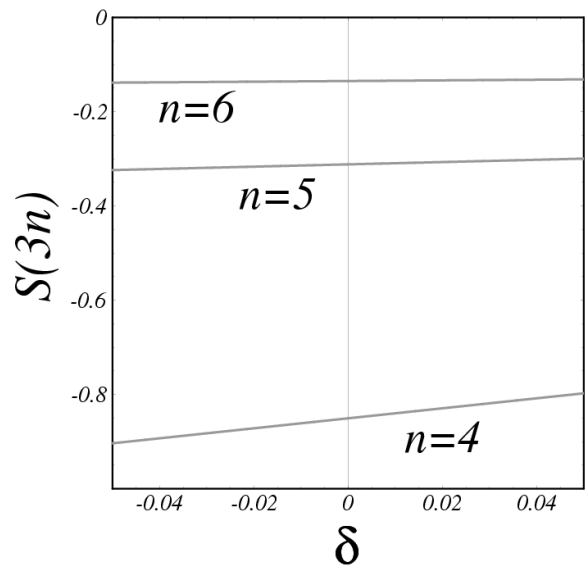}}}
      \caption{Variation of the side branching function $S(3n)=T(3n,n) + 
T(3n,2n)$ 
with the non-Newtonian parameter $\delta$, for modes $n=4,5,6$. The 
injection rate $Q=2 \pi ~{\rm cm^{2}/s}$ and $\gamma=1/100~{\rm cm^{3}/s}$. 
The units of $S(3n)$ and $\delta$ are $({\rm cm~s})^{-1}$ and $({\rm 
cm/s})^{-2}$, respectively.}
    \label{fig:5}
  \end{center}
\end{figure}

To study the growth of mode $3n$ as the non-Newtonian parameter
$\delta$ is varied, we plot $S(3n)$ as a function of $\delta$ in
Fig. 5.  We consider a particular ${\dot R} $ and $R$ combination that
corresponds to the onset of growth of mode $3n$ (i.e. obeying the
condition $\lambda(3n)=0$) in the Newtonian limit $\delta=0$. From
Fig. 5 one can verify that in the weak shear limit we consider in this
work, $S(3n)$ is negative for all values of $\delta$. As shown in
section~\ref{2nd2}, starting with a fundamental mode $a_n$, the
harmonic mode $a_{2n}$ is driven negative. Hence the product
$S(3n)a_na_{2n}$ in Eq.~(\ref{a_3n_ord2}) is positive, driving $a_{3n}>0$.

Whether side branching actually occurs depends on the magnitude of
$a_{3n}$.  Indeed, our discussion shows the presence of $a_{3n}>0$
even in the case of Newtonian fluids, where side branching does not
occur.  Why does side branching occur for shear thinning fluids? We
speculate its origin may lie in the cubic form of our modified Darcy's
law, Eq.~(\ref{new_Darcy}), which will add a term of the form $\delta
a_n^3$ to $\dot{a}_{3n}$. Unfortunately, we have not calculated this
term since we stopped our derivation at second order. Third order 
calculations for non-Newtonian Hele-Shaw flows lead to several new 
and complex mode-coupling terms, whose analysis and interpretation 
are not at all obvious, and go beyond the scope of our current work

Furthermore, there is a delicate interplay of mode $3n$ and mode $2n$,
described by the evolution equation~(\ref{a_3n_ord2}) and a similar
expression for the growth of mode $2n$,
\begin{equation}
\label{a_2n_ord2}
\dot{a}_{2n}=\lambda(2n)~a_{2n} + \frac{1}{2}~[ T(2n,n)~a_{n}^{2} + 
S(2n)~a_{n} a_{3n}],
\end{equation}
where $S(2n)=[T(2n,-n) + T(2n,3n)] >0$. Hence, side branching via a
positive $a_{3n}$ will tend to drive $a_{2n}$ positive (or at least
less negative), inhibiting tip splitting but also reducing the growth rate
of $a_{3n}$ itself.

\subsection{Rectangular geometry limit}
\label{rectangular}

It is interesting to study how the main interfacial features 
examined in the radial geometry behave if we consider flow of non-Newtonian 
fluids in {\it rectangular} Hele-Shaw cells. A mode-coupling equation 
describing the system can be obtained if we take the ``rectangular geometry 
limit'': $R \rightarrow \infty$ and $Q \rightarrow \infty$, such that $Q/(2 
\pi R) \equiv v_{\infty}$ and $n/R \equiv k$ remain constant, where 
$v_{\infty}$ is the flow velocity at infinity 
and $k$ denotes the wave number of the disturbance. In this limit the 
interface evolution reverts to the evolution of the rectangular flow 
geometry, with linear growth rate given by
\begin{equation} 
\label{RL_lambda} 
\lambda(k)=|k|\left[v_{\infty}(1+\delta v_{\infty}^{2})  
- \gamma |k|^{2}(1+2 \delta v_{\infty}^{2}) \right]. 
\end{equation} 
In addition, the only nonzero second order mode-coupling terms are
\begin{equation} 
\label{RL_GN} 
G_{N}(k,k')=|k|\left[1-{\rm sgn}(k~k^{'}) \right], 
\end{equation} 
 
\begin{equation} 
\label{RL_GNN} 
G_{NN}(k,k')=-|k|v_{\infty}^{2}\left[1-{\rm sgn}(k~k^{'}) \right], 
\end{equation} 
and 
\begin{equation} 
\label{RL_JNN} 
J_{NN}(k,k')=\frac {v_{\infty}}{2}\left \{ 
3-{\rm sgn}[k^{'}(k-k^{'})]+2~{\rm sgn}[k(k-k^{'})] \right \}. 
\end{equation} 
Note that all rectangular limit expressions~(\ref{RL_lambda})-(\ref{RL_JNN}) 
are time-independent, and preserve the reflection symmetries shown in 
Eq.~(\ref{symm}). Based on these findings, and using Eq.~(\ref{evol_eq})  we 
obtain a more compact mode-coupling equation for non-Newtonian flow in 
rectangular Hele-Shaw cells
\begin{eqnarray} 
\label{RL_2_ord} 
{\dot \zeta_{k}} &=& \lambda(k) ~\zeta_{k} + \sum_{k^{'} \neq k,0} \left 
[G_{N}(k,k^{'}) + \delta  ~G_{NN}(k,k^{'}) \right ] ~\dot 
\zeta_{k^{'}}\zeta_{k - k^{'}} \nonumber \\
                 &+& \delta \sum_{k^{'} \neq k,0}  J_{NN}(k,k^{'}) ~\dot 
\zeta_{k^{'}} \dot \zeta_{k - k^{'}}. 
\end{eqnarray}

The rectangular mode-coupling Eq.~(\ref{RL_2_ord}) is useful to 
study the influence of weak shear effects on the development of 
the fluid-fluid interface in rectangular cells. First, we examine 
finger tip-splitting related issues (finger narrowing/widening). 
As discussed in section~\ref{2nd2}, we analyze 
the influence of the fundamental $k$ on the growth of its harmonic $2k$. 
The equation of motion for the cosine mode $2k$ is
\begin{equation} 
\label{2_harm_RL} 
\dot{a}_{2k}=\lambda(2k)~a_{2k} + \frac{1}{2} ~T(2k,k) ~a_{k}^2, 
\end{equation} 
where 
\begin{equation} 
\label{T_2k_k_RL} 
T(2k,k)=\delta \left[2k^{2}v_{\infty} (v_{\infty}-\gamma ~k^{2})^{2}\right]. 
\end{equation}
Note that the driving force term $T(2k,k)$ vanishes in the Newtonian
limit $\delta=0$. This agrees with the results obtained in
Ref.~\cite{MW_rectang} for Newtonian rectangular flow. Notice further,
that for the non-Newtonian case ($\delta \ne 0$) the harmonic mode
$2k$ grows spontaneously even if missing from the initial
conditions. The selected sign of $a_{2k}$ is given by the sign of
$T(2k,k)$ and thus is dictated by $\delta$: if $\delta > 0$ ($\delta <
0$), $T(2k,k)$ is positive (negative), and hence $a_{2k}>0$
($a_{2k}<0$).

Considering the case $\delta > 0$, this means that the fingers become
narrower as the shear thinning behavior becomes more
pronounced. Moreover, by inspecting Eq.~(\ref{T_2k_k_RL}) we observe
that the width of the finger tips decreases for increasingly larger
values of the flow velocity $v_{\infty}$. Of course, these effects are
minimized for larger values of the surface tension parameter
$\gamma$. All these results are in agreement with recent
experimental~\cite{Lindner} and theoretical
investigations~\cite{Ben1,Ben2} for weak shear-thinning flows in
rectangular geometry, in which shear-thinning narrows the width of
steady state fingers.

If we take $k=k^{\star}\approx\sqrt{v_{\infty}/3\gamma}$ as the
fastest growing mode, then $\lambda(2k)<0$ so exponential growth of
$a_{2k}$ is prevented. Unlike the case of radial flow where any mode
$n$ eventually goes unstable for sufficiently large $R$, the growth
rates $\lambda(k)$ in the rectangular geometry are time
independent. Instead of exponential growth of $a_{2k}$, we instead
expect its amplitude to saturate at the value $T(2k,k) a_k^2/(-2
\lambda(2k))$ obtained by setting $\dot{a}_{2k}=0$ in
Eq.~(\ref{2_harm_RL}).  Alternatively, one could specially prepare an
initial condition with an initial perturbation of wavevector $k$
sufficiently small that $2k$ still lies within the band of unstable
modes (up to about $\sqrt{v_{\infty}/\gamma}$), in which case mode $2k$
will be spontaneously generated and able to grow to a considerable magnitude.

For completeness, we also discuss the shear thickening case: for
$\delta < 0$ the function $T(2k,k)$ is negative, favoring finger
tip-widening. Even though this shear thickening behavior has not been
studied experimentally, the broadening of the fingers in shear
thickening flow in rectangular cells has been theoretically predicted
by numerical simulations~\cite{Ben2}. Our analytical results
reinforce the correctness of such numerical predictions.

Finally, we briefly discuss side branching behavior in rectangular 
Hele-Shaw cells. We consider the influence of modes $k$ and $2k$ on the 
evolution of mode $3k$. In this case, the relevant equation of motion 
has the form
\begin{equation} 
\label{3_harm_RL} 
\dot{a}_{3k}=\lambda(3k)~a_{3k} + \frac{1}{2} ~S(3k) ~a_{k}a_{2k}, 
\end{equation} 
where, as it was in the radial flow case, the coefficient $S(3k)$ 
is the sum of $T(3k,k)$ and $T(3k,2k)$
\begin{equation} 
\label{T_3k_2k_k_RL} 
S(3k)=\delta \left[8k^{2}v_{\infty} (v_{\infty}-\gamma 
~k^{2})~(v_{\infty} - 
4\gamma k^{2}) \right]. 
\end{equation}
The favored sign of $a_{3k}$ depends on the sign of the product $a_k
a_{2k}$ and also on the sign of $S(3k)$. If $\delta$ is positive
(shear thinning) then $S(3k)$ is negative in the band of wavevectors
\begin{equation}
\label{k-band}
\sqrt{v_{\infty}/4\gamma}<k<\sqrt{v_{\infty}/\gamma}
\end{equation}
and positive outside this band. As we saw previously in the case of
mode $2k$, if the fundamental mode is taken as the fastest growing
mode $k=k^{\star}$, then exponential growth of the harmonic $3k$ is
inhibited because $\lambda(3k)<0$. Instead it will saturate at a
magnitude of $S(3k)a_ka_{2k}/(-2\lambda(3k))$.

However, for a specially prepared initial condition perturbed at
wavevector $k\approx k^{\star}/3$, the harmonics $2k$ and $3k$ will
both be able to grow.  By Eq.~(\ref{2_harm_RL}) a positive harmonic
mode $a_{2k}>0$ will appear spontaneously. Then, by
Eq.~(\ref{3_harm_RL}), modes $a_k$ and $a_{2k}$ will conspire to
create mode $a_{3k}$. Because $k\approx k^{\star}/3$ lies {\em outside}
the wavevector band~(\ref{k-band}), the value of $S(3k)$ is {\em
positive}. Hence the mode $a_{3k}$ is driven positive, the sign needed
to create side branches. While this conclusion is only tentative (due
to our neglect of the complete set of third order terms) it would be
interesting to test via numerical simulation or experiment.

\section{Concluding remarks}
\label{conclude}
Visually striking patterns arise when a less viscous Newtonian fluid
displaces a more viscous non-Newtonian fluid in the confined geometry
of a Hele-Shaw cell. These complex patterned structures are the result
of finger tip-splitting and dendritic interfacial instabilities. Due
to the complicated dynamics of the system, and also from the
limitations imposed by purely linear analysis, the majority of the
theoretical studies in this area of research rely heavily on
sophisticated numerical methods. In this work, we developed an
alternative theoretical approach to the problem which allowed us to
address important nonlinear issues analytically.  A key point in our
derivation was the introduction of velocity vector potential (as
opposed to scalar potential) which was demanded by the non-Newtonian
fluid flow.

We started our investigation by analyzing the role of fluid viscosity
anisotropy on the development of the Saffman-Taylor instability in
radial Hele-Shaw cells.  To approach the problem analytically we
considered the weak shear limit, and focused on the onset of nonlinear
effects. In order to examine the influence of shear thinning and shear
thickening on the shape of the emerging patterns, we derived a
mode-coupling equation which is ideally suited to describe the weakly
nonlinear interface evolution. Our analytical results show that finger
tip-splitting is enhanced (diminished) in the case of shear thickening
(thinning) fluids. We applied the rectangular geometry limit to our
radial mode-coupling equations of motion, and also studied finger-tip
and side branching behavior in rectangular cells. Our main findings
predict finger tip-widening (narrowing) for shear thickening
(thinning) fluids.

In neither the radial nor the rectangular geometry limit were we able
to show the spontaneous appearance of side branching, except for some
specially prepared initial conditions in rectangular geometry. We
speculate that extending the current approach to third order could
shed more light on this problem.

In summary, our analytical mode-coupling approach detects and predicts 
several features of the patterns formed in non-Newtonian Hele-Shaw flows 
in both radial and rectangular geometries.
Furthermore, it predicts some behaviors not yet investigated in the 
literature, such as the evolution mechanisms involving tendency towards 
advanced finger tip-splitting and reduction of side branching for shear 
thickening fluids. We hope the main results presented in this paper 
will prompt further theoretical and experimental work 
on non-Newtonian Hele-Shaw flows, especially in the area of early stage
transients in rectangular geometry where our theory makes specific
predictions that have not been subject to experimental or numerical check.

\begin{acknowledgments}

J.A.M. thanks the Brazilian Research Council - CNPq (through its PRONEX 
Program) for financial support. Work of M.W. was supported in part by the 
National Science Foundation grant No. DMR-0111198. M. C. acknowledges support 
from Carnegie Mellon University and University of Maryland (National Science 
Foundation grant No. DMR-00-80008).

\end{acknowledgments}

\appendix*
\section{Second order mode-coupling terms}
\label{app}

This appendix presents the expressions for the second order 
Newtonian (N), and non-Newtonian (NN) mode-coupling coefficients which appear 
in Eq.~(\ref{2_ord_eq}) 
\begin{equation}
\label{F_N}
{F_{N}\left (n,m \right)} = \frac{|n|}{R} \left \{ \frac{{\dot R}}{R} \left[ 
\frac{1}{2}-{\rm sgn}(nm) \right] -
\frac{ \gamma}{R^{3}} \left(1- \frac{nm}{2}- \frac{3m^{2}}{2} \right) \right 
\},
\end{equation}  

\begin{eqnarray}
\label{F_NN}
F_{NN}(n,m)& = &\frac{|n|}{R} {\dot R}^{2} \Bigg \{ 
\frac{{\dot R}}{R} \left [ \frac{2 \alpha(m)}{m}\frac{|n|}{n} - 
\frac{2 \beta(n)}{|n|} +  \frac{f(n,m)}{2(|n|+1)} \right ] \nonumber\\
           & - &\frac{\gamma}{R^{3}} \frac{2|n|}{(|n|+1)} \left ( 
1-\frac{nm}{2} - \frac{3m^{2}}{2} \right ) \Bigg \},
\end{eqnarray}  

\begin{equation}
\label{G_N}
G_{N}\left (n,m \right)= \frac{|n|}{R} \left[1-{\rm sgn}(nm) -\frac{1}{|n|} 
\right],
\end{equation} 

\begin{equation}
\label{G_NN}
G_{NN}\left (n,m \right)= \frac{|n|}{R} {\dot R}^{2} \left [ \frac{2 
\alpha(m)}{m}\frac{|n|}{n} - 
\frac{2 \beta(n)}{|n|} +  \frac{f(n,m)}{2(|n|+1)}\right ],
\end{equation}   
 
\begin{equation}
\label{H_NN}
H_{NN}\left (n,m \right)= \frac{|n|}{2(|n|+1)} \frac{{\dot R}^{2}}{R}~f(n,m),
\end{equation}
and
\begin{equation}
\label{J_NN}
J_{NN}\left (n,m \right)= \frac{|n|}{2(|n|+1)} {\dot R} ~f(n,m),
\end{equation}   
where
\begin{equation}
\label{f}
f(n,m)=\left[ \frac{|n|}{n} h \left(n,m \right) + 
k \left(n,m \right) \right].
\end{equation}

\end{document}